# DEVELOPMENT AND APPLICATION OF A DIAPHRAGM MICRO-PUMP WITH PIEZOELECTRIC DEVICE

*H.K. Ma[1], B. R. Hou[1], H. Y. Wu[1], C.Y. Lin[1], J. J. Gao[1], M.C. Kou[2]*

Department of Mechanical Engineering[1]
National Taiwan University, Taipei, Taiwan
Cooler Master Co., Taiwan[2]
skma@ntu.edu.tw, 8862-23629976

**ABSTRACT**

In this study, a new type of thin, compact, and light weighed diaphragm micro-pump has been successfully developed to actuate the liquid by the vibration of a diaphragm. The micro-diaphragm pump with two valves is fabricated in an aluminum case by using highly accurate CNC machine, and the cross-section dimension is 5mm×28mm. Both valves and diaphragm are manufactured from PDMS. The amplitude of vibration by a piezoelectric device produces an oscillating flow which may change the chamber volume by changing the curvature of a diaphragm. Several experimental set-ups for performance test in a single micro-diaphragm pump, isothermal flow open system, and a closed liquid cooling system is designed and implemented. The performance of one-side actuating micro-diaphragm pump is affected by the design of check valves, diaphragm, piezoelectric device, chamber volume, input voltage and frequency. The measured maximum flow rate of present design is 72 ml/min at zero total pump head in the range of operation frequency 70-180 Hz.

## 1. INTRODUCTION

Micro-pumps have been developed by using several actuation methods, such as electromagnetic [1], piezoelectric [2-7], shape memory alloy [8], electrostatic [10], and thermo-pneumatic [11] devices. Most of them have complex structures and lower pumping rates. On the contrary, piezoelectric actuation has advantages of the relatively simple structure and lower power consumption. Micro-diaphragm pump can be classified as with or without valve [2, 9-13]. A valve-less pump consists of two fluid flow rectifying diffuser/nozzle elements which are connected to the inlet and outlet of a pump chamber with a flexible diaphragm. Stemme [6] proposed the first prototype valveless pump consisting of a circular cylindrical volume where the top side had a thin brass diaphragm to which a piezoelectric disc was fixed. Its flow rate was 15.6 ml/min. Olsson et al. [7-9] investigated the flow-directing properties of several diffuser elements with different lengths and opening angles for valveless micro-pumps. Numerical simulations were done by using the Computational Fluid Dynamics program ANSYS/Flotran. Our study group also designed and simulated a valveless micro-pump which has a 0.5 mm thick pump chamber and a 20 degree open angle [14]. The peak flow rate in the simulation was only 0.12 ml/min. In general, a valveless pump can't provide a higher flow rate in a closed system with higher flow resistances. Therefore, a new design of one-side actuating micro-diaphragm pump is developed to actuate the liquid by the vibration of a diaphragm with a thinner structure and a higher flow rate.

## 2. DEVELOPMENT OF A PIEZOELECTRIC MICRO-DIAPHRAGM PUMP

Generally, a diaphragm pump with passive check valves can overcome the higher flow resistance and improve the performance. Ordinary micro-pumps cannot provide enough flow rates due to the limited actuating force by diaphragm's displacement. Böhm [1] developed a plastic micro-pump with a valve which is capable of pumping both liquid and gas at pump rates 2 ml/min for water and up to 50 ml/min for air when actuation frequencies between 2 and 500 Hz. A commercial product of SDMP305D with central-actuating [15], which is shown in Fig. 1, is a thin, compact, and lightweight micro-diaphragm pump. However, it is limited to the height, the control volume, and the mass flow rate due to its actuating displacement in z-direction by Piezoelectric effective. Its maximum flow rate is only 5 ml/min.

### 2.1. One-side Actuating Micro-diaphragm Pump

In Fig. 3, a new micro-diaphragm pump with piezoelectric effect has been designed to actuate the working fluid by the vibration of a diaphragm with one-side sector-shaped piezoelectric device. The vibration amplitude of a piezoelectric device produces an oscillating flow which may cause the chamber volume change by altering the curvature of a diaphragm. While the actuator is moving downward for decreasing chamber volume shown in Fig. 2(a), the outflow will be in one direction with closed inlet valve and opened outlet valve. While the actuator is moving upward to increase chamber volume shown in Fig. 2(b), the inflow will be in the chamber with the inlet valve open and the outlet valve closed. The new design of the one-side actuating diaphragm pump with two check valves, which is shown in Fig. 3, can allow the pump thinner and flow in one direction. In addition, the actuating force can be enforced by its harmonic resonance of working fluid with the vibration of a rectangular piezoelectric device, PDMS diaphragm, and two check valves in the pump chamber.

The micro-diaphragm pump with two valves is fabricated in an aluminum case, and the cross-section dimension is 5mm×28mm (without cover). Both check valves and diaphragm are manufactured from PDMS.

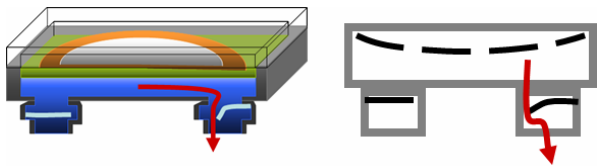

**Figure 1:** View of SDMP305D micro-diaphragm pump

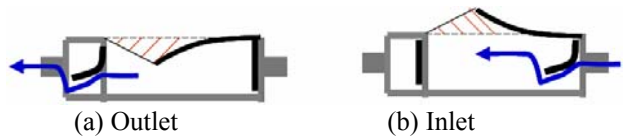

(a) Outlet          (b) Inlet
**Figure 2:** One-side actuating micro-diaphragm pump

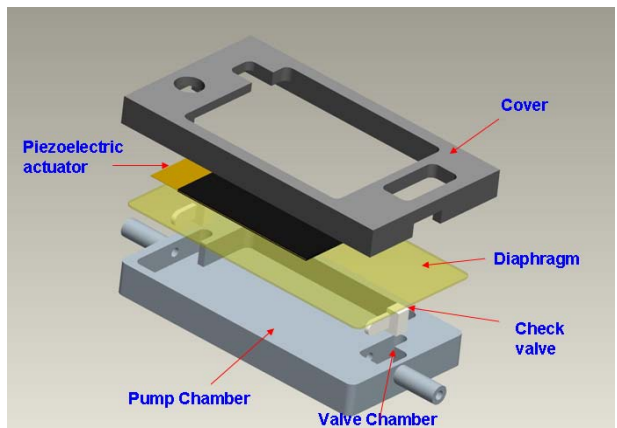

**Figure 3:** Newly designed micro-diaphragm pump

**2.2. Theoretical Analysis**

The passive check valve is an important device in a new design of one-side actuating micro-diaphragm pump. It decides the performance of the pump. The governing equation of oscillating motion for valves, which is shown in Fig. 4, can be expressed by Eqn. (1)

$$m\frac{d^2 y}{dt^2} + c\frac{dy}{dt} + ky = F\sin(\omega \cdot t) \qquad (1)$$

Where $m, c, k$ are mass, damping coefficient, and spring constant, respectively. The right-hand term, $F \cdot sin(\omega \cdot t)$, represents the external force from actuating diaphragm acting on the valve. $\omega = 2\pi f$ is the frequency of the piezoelectric device. The left-hand side of the equation describes the force from the system components that comprise a single degree-of-freedom system. The inertia force of valve with mass m is the change rate of linear momentum m·a. The damping force generated by the viscous fluid is proportional to the pressure drag and friction drag. The damping factor $\zeta$ is defined as

$$\zeta = \frac{c}{2\sqrt{k \cdot m}} \qquad (2)$$

The damping factor is a function of the shape of the valve, the kinematic viscosity of the fluid, and the frequency of oscillation of the valve. Check valves are manufactured from PDMS. It is a flexible, transparent elastomer ideally suited for electrical/electronic potting and encapsulating applications. The elastomer is elastic structural element which can be expressed as spring motion to store and release the energy. In this study, the shape of the check valve can be considered as a cantilever beam. The bending moment is a function of position along the valve, and the deformation of the valve is small. Thus, the differential equation for the elastic curve of the valve can be shown as:

$$EI\frac{d^2 y}{dx^2} = M(x) \qquad (3)$$

Where $E$ is the module of elasticity; $I$ is the second moment of area of the transverse section with respect to the neutral axis; and $M$ is the bending moment of the section. The spring constant $k$ of the cantilever valve can be calculated from Eqn. (3) as:

$$k = \frac{3EI}{L^3} \qquad (4)$$

Where $L$ is the length of the valve, and $I = bh^3/12$ is the second moment of area of a rectangular section.
For translation oscillations of the valve, the natural frequency $\omega_n$ of the valve in vacuum is defined as:

$$\omega_n = \sqrt{k/m} \qquad (5)$$

**2.3 Head Loss**

For steady, inviscid, imcompressible flow the total energy remains constant along a streamline. The energy line also represents the total head available to the fluid. For considering the friction drag, the total head loss $H_L$ can be calculated by the Bernoulli equation and be written as

$$\frac{P_1}{\rho g} + \frac{1}{2}\frac{V_1^2}{g} + z_1 = \frac{P_2}{\rho g} + \frac{1}{2}\frac{V_2^2}{g} + z_2 + H_L \quad (6)$$

Both the pressure of the inlet $P_1$ and the outlet $P_2$ are the atmospheric pressure. The total head loss may further be reduced by neglecting the inlet velocity $V_1$ and setting $Z_2$ as zero.

$$H_L = z_1 - \frac{1}{2}\frac{V_2^2}{g} \quad (7)$$

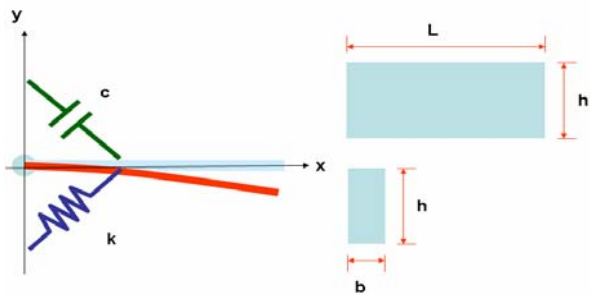

**Figure 4:** A single degree-of-freedom system

## 3. EXPERIMENTAL SET-UP

Fig. 5 shows the experimental setup for the performance test of micro-diaphragm pump. The micro-diaphragm pump with piezoelectric device is driven by an alternating sine-wave voltage of ±50V at 70-180 Hz from a function-generator and an amplifier. The signal waves and frequencies are controlled by a function-generator. An amplifier is needed to magnify voltage to designate dB, and input signals can be monitored by an oscilloscope. The water flow rate data are recorded every 10Hz to analyze the performance of the pump in different potential heights and operating conditions.

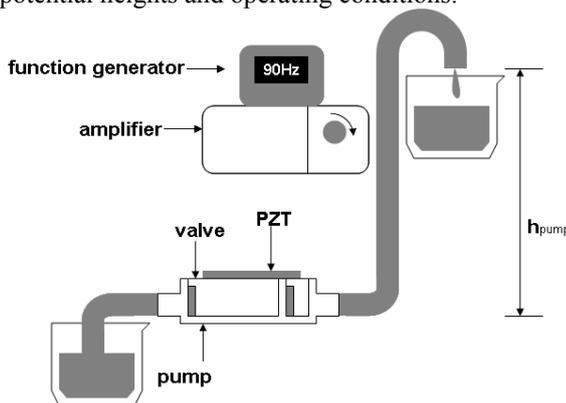

**Figure 5:** Set-up for pump performance

Without activating the pump, certain flow velocities through the whole system are obtained by adjusting the water level. As shown in Fig. 6, the resistance in the pump ($H_{pump}$) can be calculated by the total head loss of the whole system ($H_L$) minus the head losses from $H_{coldplate}$ and $H_{pipe+other}$. The resistance in the cold plate ($H_{coldplate}$) can be measured in a similar way.

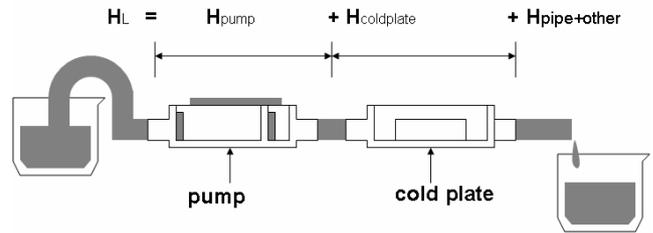

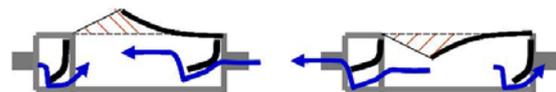

**Figure 6:** Set-up for resistance measurement

## 4. RESULTS AND DISCUSSION

### 4.1. Effects of Actuating Frequency on Flow Rates

The flow rate gradually rises as frequency increases in a well designed pump. However, most of the flow rate-frequency curves (Fig.9 & 10) show two or more peak values at different frequencies in different designs. Very low flow rate at a higher frequency may occur in bad designs. This is enormously relevant to the vibration amplitude of a piezoelectric device which may produce oscillating flows by changing the curvature of a diaphragm. The flow rate drops correspondingly if the valve does not function properly, as the actuator moves downwards and upwards in order to change chamber volume. In an abnormal actuating operation shown in Fig. 7, the outlet valve may not be able to avoid back flow as the actuator vibrates at a certain frequency as the control volume changes. The motion of the outlet valve mismatches the vibration of membrane may cause the inability to produce a net outflow.

**Figure 7:** Phenomena in abnormal actuating cases

### 4.2. Effect of diaphragm Valve Thickness

In general, the oscillating flow generated by the vibration amplitude of a piezoelectric device can be affected by changing the thickness of a diaphragm. A thinner diaphragm will deform unpredictably beside the actuator. Consequently, the change of the control volume decreases or even becomes zero. This causes the deactivation of the pump. On the contrary, a thicker membrane has larger tension. The actuator has to overcome the tension of the membrane as well as to drive the working fluid in the pump. In this study, the dimension of a standard valve is 5mm×3mm, and the standard diaphragm is 0.03g/cm$^2$.

Fig. 8 is the comparison of measured flow rate profiles obtained under different valve thicknesses. Obviously valve thickness has a big effect on the flow rate. The higher flow rates may occur at frequency 130 Hz and 180 Hz by using 0.5mm valve thickness. The resonance phenomena happened due to the vibration of working fluid with system components (valve, diaphragm) in the pump chamber. The resonance frequency determines where the highest flow rate locates. Usually the valve is constrained by viscosity and inertia force caused by the working fluid. Under a fixed inertia force, the thicker the valve is, the smaller the response acceleration will be; that is, the smaller displacement will be made. The small gap between the outlet pipe and the valve causes high resistance in flow and thus results in reducing the flow rate. Similar phenomena will exist in different valve thicknesses.

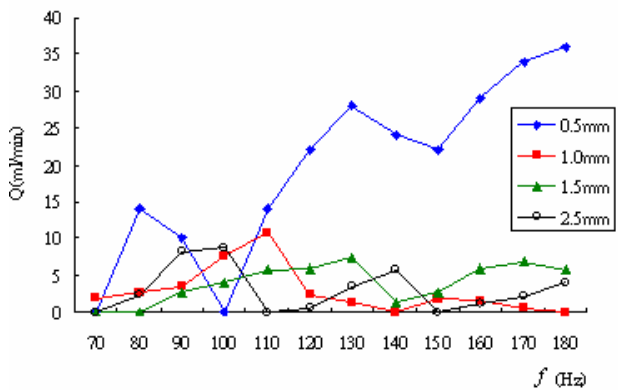

**Figure 8:** Measured flow rate with different valve thicknesses under different frequencies (at ±50V)

### 4.3. Effect of Valve Length and Shape

Fig.9 is the comparison of measured mass flow rate profiles obtained under different types of valve. Obviously the valve length and shape also have large effects on the mass flow rate. The short or narrow-type valve may have a higher flow rate at the resonance frequency due to a smaller mass and a lower drag resistance. The drag coefficient of a valve depends on the viscosity of the fluid and the shape of the valve. In addition, the deflection of the piezoelectric device depends on the input voltage. It also affects the flow rate directly, as shown in Fig. 10.

### 4.4. Pump Performance under Certain Pressure Heads

Fig. 11 shows the pump performance at various pressure heads. The pressure head represents the distance between the water surface and the pump. The flow rate varies with the height of the pipe outlet as shown in Fig. 5. It approaches to zero at 0.52 m higher to the pump. The relation of flow rate with pressure heads is linear.

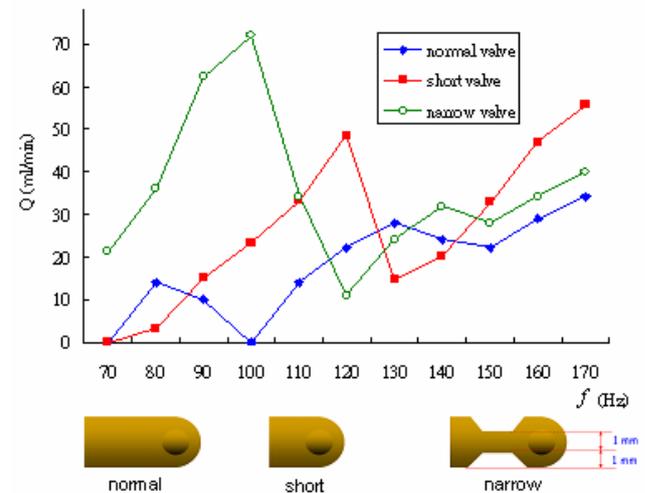

**Figure 9:** Measured flow rate with different valve type under different frequencies (0.5mm thickness at 50V)

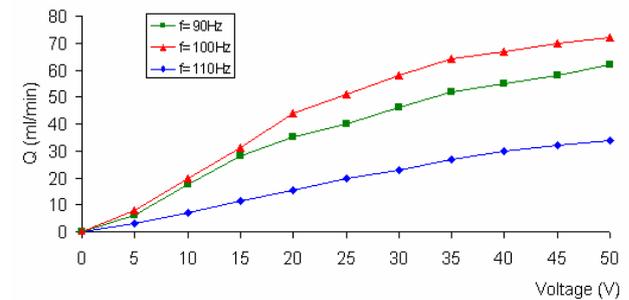

**Figure 10:** Measured flow rates at different voltages (with 0.5mm thick narrow valves)

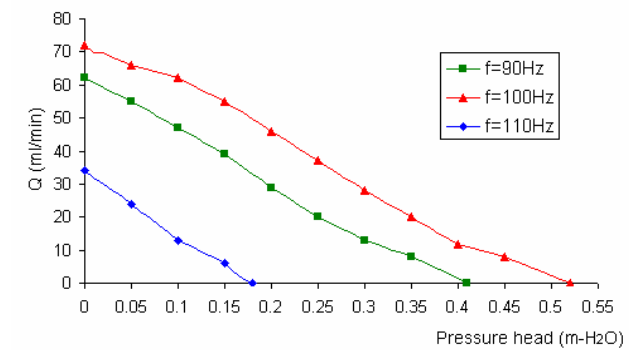

**Figure 11:** P-Q curves of the micro-pump (with 0.5 mm thick narrow valves)

### 4.5 Head Losses in a Closed System

Fig. 12 shows the head losses in system, pump and cold plate linearly increase with the flow velocity. The resistance in pump is the major head loss in whole system. The results also show the fact that the head loss is proportional to the flow velocity due to one-side micro-diaphragm pump has higher power to actuate water through the closed system.

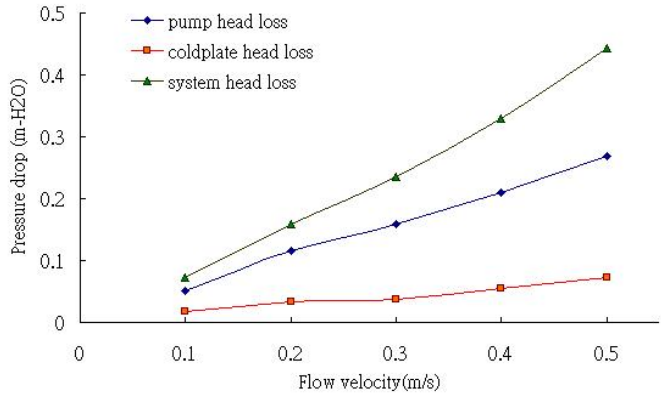

**Figure 12:** Head losses of components (0.5mm thickness at 50V)

## 5. APPLICATIONS OF A PIEZOELECTRIC MICRO-DIAPHRAGM DEVICE

In recent years, temperature control of electronic devices has become more important because heat dissipation has increased substantially, due to the device miniaturization, increased heat flux in gaming notebook computer (NB), and workstation NB. Increasing temperature directly affects device's performance and reliability. Therefore, the thermal management plays a vital role for Central Processing Unit (CPU) and Graphics Processing Unit (GPU) in NB design.

### 5.1 Application of Micro-diaphragm Pump in a Closed Water Cooling NB System

This study focus on performance of a micro-diaphragm pump applied in a closed water cooling system, by simple designs of a cold plate and a radiator in the testing system. The relation between the heat generated in a CPU and its core temperature is considered. The measured results (Fig. 13) show that the designed cooling system keeps a core temperature in CPU at 48°C during a 30W operation. It is a sufficient resolution to the heat releasing of present laptops on full speed operation. The results also show the CPU core temperature rises as its operation power increases. When the CPU power reaches 60W, the core temperature is 73.6°C.

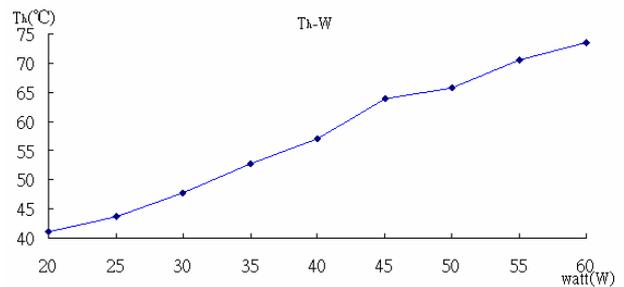

**Figure 13:** The performance of water cooling system under different frequencies in a close loop (0.5mm thickness narrow valve)

### 5.2 Application in Fuel Cell System

We may apply piezoelectric effect to construct micro-diaphragm pumps as the oxygen feeding system for novel micro PEM fuel cells. The flow of hydrogen is driven by a hydrogen tank at the anode. While the actuator is moving outward/inward for expanding/compressing chamber volume shown in Fig. 14 and the air (contains the oxygen) will be sucked or compressed into the catalyst layer due to diaphragm vibration. Also, the product of water will be removed from the catalyst layer. Moreover, the new design of micro-diaphragm flow channel with piezoelectric effect can make the flow turbulent which increase the chemical reaction and thus increase current density.

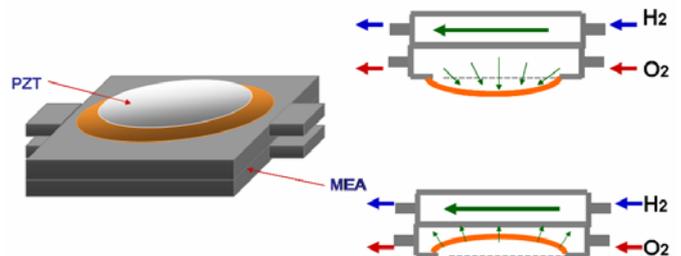

**Figure 14:** Application in Fuel Cell System

## 6. CONCLUSIONS

A new design of one-side actuating micro-diaphragm pump with piezoelectric device has been successfully developed by making use of its harmonic resonance of working fluid with system components (valve, diaphragm) in the pump chamber. Its performance is affected by the design of check valves, diaphragm, piezoelectric device, chamber volume, voltage and frequency. It shows the following conclusions:
1. The measured maximum flow rate of new design pump is 72 ml/min at zero pump head in the range of operation frequency 70-180 Hz

2. Narrow valve with 0.5 mm thickness is an optimal value for higher pump flow rate, and the maximum voltage of a piezoelectric device is ±50V (AC).
3. The head of the micro-pump with narrow valves can reach to 0.52 m.
4. The newly designed micro-pump will be widely tested with the advantages of thin size, powerful flow rate, less leaking problem, and durability in the fields of closed system, heat dissipation, and fuel cell system in the future.

**Acknowledgments**

The authors thank the Cooler Master Co. and NSC (NSC94-2621-Z-002-007) for supporting this project.